\newcommand{\dfd}{{\rm d}}
\newcommand{\vecr}{\bm{r}}
\newcommand{\vecs}{\bm{s}}
\newcommand{\vecv}{\bm{v}}
\newcommand{\vece}{\bm{e}}
\newcommand{\vecA}{\bm{A}}
\newcommand{\vecB}{\bm{B}}
\newcommand{\vecOmega}{\bm{\Omega}}
\newcommand{\ZF}{Z_{\mbox{\tiny F}}}
\newcommand{\EF}{{\cal E}_{\mbox{\tiny F}}}
\newcommand{\vF}{v_{\mbox{\tiny F}}}

\documentclass[aps,pre,groupedaddress,endfloats,showpacs]{revtex4}
\usepackage{bm}
\begin{document}
\title{Circulating electrons, superconductivity, and the
Darwin-Breit interaction}
\author{Hanno Ess\'en}
\affiliation{Department of Mechanics \\Royal Institute
of Technology  \\ S-100 44 Stockholm, Sweden}
\date{October 2013}
\begin{abstract}
The importance of the Darwin-Breit interaction between electrons
in solids at low temperatures is investigated. The model
problem of particles on a circle is used and applied to
mesoscopic metal rings in their normal state. The London moment
formula for a rotating superconducting sphere is used to calculate
the number, $N$, of superconducting electrons on the sphere. This
number is found to be three times the radius, $R$, of the sphere
divided by the classical electron radius, i.e.\ $N=3R/r_{\rm e}$.
The Darwin-Breit interaction gives a natural explanation for this
relation from first principles. It also is capable of electron
pairing. Collective effects of this interaction require a
minimum of two dimensions but electron pairing is enhanced in
one-dimensional systems.
\end{abstract}
\maketitle



\section{Introduction}
Arguments and results will be presented that hopefully convince
the open-minded reader that superconductivity is caused by the
Darwin-Breit (magnetic) interaction between semiclassical
electrons. The starting point is a careful study of the model
problem of electrons on a circle. This simple model is chosen
since it allows accurate treatment of the notoriously difficult
problem of relativistic and magnetic effects in many-electron
systems. Since classical ideas are closer to our intuition the
classical picture is taken as far as possible before quantum
mechanics is reluctantly adopted.  The semiclassical point of
view is an extremely powerful one
\cite{BKbrack,BKgutzwiller} and the reader will find further
examples of this below.

Relativistic quantities, to a first approximation,  have a
magnitude $(v/c)^2$ times those of non-relativistic quantities.
While this always is small in everyday life, in the atomic world
this parameter is $\sim 10^{-4}$, which is fairly small, but
rarely negligible. A striking example of this is the energy gap
in superconductors which typically is order of magnitude
$10^{-4}$ of the Fermi energy. Any study of this phenomenon that
does not take relativistic effects into account must consequently
remain inconclusive

The Darwin-Breit interaction \cite{darwin,breit29,breit32} is the
first order relativistic  correction,
\begin{equation}
\label{eq.V1.velocity.form}
V_1=-\sum_{i<j}^N \frac{e^2}{c^2}\frac{\vecv_i\cdot\vecv_j
+ (\vecv_i\cdot\vece_{ij}) (\vecv_j\cdot\vece_{ij})}{2 r_{ij} },
\end{equation}
to the Coulomb potential. Sucher \cite{sucher} in a recent review
({\em What is the force between two electrons?}) gives a
thorough discussion of its origin in QED. While well known as an
important perturbation in accurate atomic calculations
\cite{BKstrange,cook} it has until recently (Ess\'en
\cite{essen95,essen96,essen97,essen99}) usually been taken for
granted, without proof or justification, that it is negligible in
larger systems. Welker \cite{SCwelker} suggested in 1939 that
magnetic attraction of parallel currents might cause
superconductivity, but after that the idea seems to have been
forgotten. Other types of magnetic interaction have been
suggested though \cite{SCmathur}. Some efforts to include the
Darwin-Breit interaction in density functional approaches to
solids are reviewed in Strange \cite{BKstrange}. Capelle and Gross
\cite{SCcapelle} have also made efforts towards a relativistic
theory of superconductivity.

In section \ref{sec.rings.flux} we introduce the analytical
mechanics of particles on a circle and apply it to mesoscopic
rings. This serves to introduce the mathematical model and also
throws some light of the theory behind the persistent currents
found in these. We later find that, though these rings are not
superconducting, electron pairing might be relevant to understand
their physics.

Section \ref{sec.london.moment} closes in upon the main subject
of superconductivity. The London moment formula connecting the
angular velocity of a superconducting body and the magnetic
field it produces is introduced and motivated. The formula,
together with classical electromagnetism can be used to
calculate the number of superconducting electrons present. This
number is found to be determined entirely by fundamental
constants and the size of the body.

Finally in section \ref{sec.darwin.breit} the importance of the
Darwin-Breit interaction is investigated. We show how it can lead
to electron pairing and calculate the relevant temperatures at
which these form. We also investigate when the interaction
might become dominating and find that exactly the combination of
number, size, and fundamental constants that followed from the
London moment is the condition for this. When the condition is
fulfilled the particles no longer move individually, or in pairs,
but collectively. The behavior of this condition as a function
of spatial dimension is investigated. Interestingly it is found
that the one-dimensionality of the ring enhances pair-formation
but suppresses collective behavior (superconductivity).
After that the conclusions are summarized.

\section{Rings, persistent currents, and flux periodicity}
\label{sec.rings.flux}
In solid state physics cold mesoscopic metal rings have attracted
a lot of attention. In particular since theoretical predictions
\cite{SCbuttiker,SCbloch} that an external magnetic flux through the
ring causes a persistent current round it, have been
experimentally verified \cite{SCwebb,SClevy,SCchandrasekhar}. The
agreement between theory and experiment is, however, still far
from perfect \cite{SCjohnson}, for reviews see
\cite{SCimry,BKimry}. One normally assumes that it is
correct to treat the conduction electrons semiclassically, one
speaks about ballistic electrons \cite{BKbrack,BKimry},
and we will do so here. Superconductivity is not treated in this
section, but we assume that the rings are perfect conductors
(have zero resistance).

\subsection{Charged particles on a circle}
We now set up the model problem of charged particles constrained
to move on a circle. Assuming that the circle has radius $R$,
positions and velocities are given by
\begin{equation}
\label{eq.pos.vel.vectors.meso}
\vecr_i(\varphi_i)=R \vece_{\rho}(\varphi_i), \hskip 0.3cm
\mbox{and} \hskip 0.3cm
\vecv_i(\varphi_i,\dot\varphi_i)=
R \dot\varphi_i \vece_{\varphi}(\varphi_i),
\end{equation}
where  $\vece_{\rho}(\varphi)= \cos\varphi\, \vece_x
+\sin\varphi\,\vece_y$ and $\dot{\vece}_{\rho}
=\dot\varphi\vece_{\varphi}$, as usual. We take the zeroth order
Lagrangian to be
\begin{equation}
\label{eq.zeroth.circle.lagrangian.V0}
L_0=T_0-V_0=\frac{1}{2} \sum_{i=1}^N m_i R^2\dot\varphi_i^2  -
V_0(\varphi_1,\ldots,\varphi_N).
\end{equation}
Since we will have metallic conduction electrons in mind the
potential $V_0$ does not necessarily represent the Coulomb
interactions, but rather interactions with the lattice plus,
possibly, Debye screened two particle interactions. The
generalized (angular) momenta are
$J_i=\partial L_0/\partial\dot\varphi_i = mR^2\dot\varphi_i$ so
the Hamiltonian is
\begin{equation}
\label{eq.basic.ring.hamiltonian}
H_0=\sum_{i=1}^N \frac{J_i^2}{2m_i R^2} + V_0.
\end{equation}
If there is a magnetic flux $\Phi =\int \vecB\cdot\dfd\vecs
=\oint \vecA\cdot\dfd\vecr =2\pi R A_{\varphi}$ through the ring
the Hamiltonian changes to
\begin{equation}
\label{eq.ring.hamiltonian.ext.A}
H_0=\sum_{i=1}^N \frac{1}{2m_i}\left(
\frac{J_i}{R} -\frac{e_i}{c}A_{\varphi} \right)^2 + V_0=
\sum_{i=1}^N \frac{1}{2m_i R^2}\left(
J_i -\frac{e_i}{2\pi c} \Phi \right)^2 + V_0,
\end{equation}
since $A_{\varphi}=\Phi/(2\pi R)$.

We find the equations of motion
\begin{eqnarray}
\dot J_i = -\frac{\partial H_0}{\partial \varphi_i}
=-\frac{\partial V_0}{\partial \varphi_i},\\
\dot \varphi_i = \frac{\partial H_0}{\partial
J_i}=\frac{J_i}{m_i R^2}-\frac{e_i \Phi}{m_i R^2 2\pi c}.
\end{eqnarray}
The current round the ring is by definition
\begin{equation}
\label{eq.ring.current.ext.A}
I=\sum_{i=1}^N e_i \frac{\dot\varphi_i}{2\pi} =\frac{1}{2\pi}
\sum_{i=1}^N\left( \frac{e_i J_i}{m_i R^2}-\frac{e_i^2 \Phi}{m_i
R^2 2\pi c} \right) \equiv I_0 + I_{\Phi}.
\end{equation}
One notes that the relation
\begin{equation}
\label{eq.ring.current.dH.dA}
I=-c\frac{\partial H_0}{\partial \Phi}
\end{equation}
holds.

For non-interacting particles on the ring we have
\begin{equation}
V_0=\sum_{i=1}^N U_0(\varphi_i).
\end{equation}
Then $H_0=\sum_i H_i(J_i,\varphi_i)$ where $H_i$ are constants
of the motion, $H_i=E_i$, whether there is a flux or not. There are
then the adiabatic invariants \cite{BKlandau1}
\begin{equation}
I_{\varphi_{i}} \equiv \frac{1}{2\pi} \oint J_i(\varphi_i
;E_i,\Phi) \dfd\varphi_i = \overline{J_i},
\end{equation}
the averages, $\overline{J_i}$, of the $J_i$ round the ring.
If the flux is turned on slowly they will retain their zero
flux values.  The zero flux average current
\begin{equation}
\label{eq.ring.current.A.0}
\overline{ I_0} =\frac{1}{2\pi R^2} \sum_{i=1}^N \frac{e_i
\overline{J_i}}{m_i}
\end{equation}
is thus also an adiabatic invariant, and remains constant. This
means that slowly turning on a flux $\Phi$ through the ring
results in the extra diamagnetic circulating current
\begin{equation}
\label{eq.ring.current.due.A}
I_{\Phi} = -\frac{\Phi}{4\pi^2 R^2} \sum_{i=1}^N
\frac{e_i^2}{m_i c}
\end{equation}
independently of any pre-existing current. Below we will find that
the above result can be found using Larmor's theorem and thus, in
fact, is independent of electron interactions provided other
conditions are fulfilled.

\subsection{Two types of current}
We find that there are two different types of current
possible in these rings. The `ballistic' current $I_0$, which
should be, at most \cite{SCgeller}, order of magnitude a few $ e
\vF/(2\pi R)$, where $\vF$ is the Fermi velocity, and the Larmor
current $I_{\Phi}$ induced by the flux. Assuming that only
electrons contribute (\ref{eq.ring.current.due.A}) becomes
\begin{equation}
\label{eq.ring.current.due.A.elect}
I_{\Phi} = \frac{\Phi}{4\pi^2 R^2} N\frac{e^2}{mc} .
\end{equation}
Putting
\begin{equation}
\label{eq.flux.unit}
\Phi = n_{\phi} \frac{hc}{|e|} \equiv n_{\phi} \Phi_0 ,
\end{equation}
where $n_{\phi}$ is dimensionless and $\Phi_0=hc/|e|$ is the flux
quantum, we get the expression
$ I_{\Phi} \pi R^2 = - N n_{\phi} \mu_{\rm B}$.
Here   $\mu_{\rm B}=|e|\hbar/(2m)$ is the Bohr magneton. Gaussian
units are used in most formulas; to get equation
(\ref{eq.ring.current.due.A.elect}) in SI-units we simply delete
$c$. If the flux is $\Phi=B \pi R^2$ we can then rewrite it in
the form
\begin{equation}
\label{eq.ring.current.due.A.SI}
I_{\Phi} = -N\cdot B\cdot 2.242 \,{\rm nA/T}.
\end{equation}
To get a number out of this formula we must estimate the number
$N$ of semiclassical electrons and know the magnetic field in
teslas. The speed corresponding to the Larmor current is, in atomic
units, $v_{\Phi}=n_{\phi}/R \ll \vF=1.92/r_s$, where $r_s$ is the
radius parameter. On the other hand all semiclassical electrons
contribute to $I_{\Phi}$, whereas the number contributing to $I_0$
necessarily is small.

Levy et al.\ \cite{SClevy} found an average current of
$I_{\rm av}=3\cdot 10^{-3} \cdot e \vF/\ell= 0.36\,$nA in their
Cu-rings, of circumference $\ell=2.2\,\mu$m. If this is
interpreted as a Larmor-current we can calculate $N$. At the
magnetic field $B_0=1.3 \cdot 10^{-2}\,$T corresponding to the
flux quantum $\Phi_0$ this gives the  reasonable result $N \approx
100$ for the number of semiclassical electrons in the system.
Chandrasekhar et al.\ \cite{SCchandrasekhar}, on the other hand,
found currents $I=$(0.3 -- 2.0) $e\vF/(2\pi R)$ in a single gold
ring. These can thus only be interpreted as due to ballistic
currents. They might be due to electron pairs, which may form
even in the normal state, as we will see below.

\subsection{Larmor's theorem}
Consider a system of particles, all of the same charge to mass
ratio $e/m$. Assume that they move in a common external potential,
$U_e(\rho,z)$, that is axially symmetric, i.e.\ independent of
$\varphi$, under the influence of arbitrary interparticle
interactions. Now place this system in a weak magnetic
field, $B_z$, along the $z$-axis. One can then apply Larmor's
theorem \cite{BKstrange,essen89} to show that the response of the
system to this field is a rotation with angular velocity
\begin{equation}
\label{eq.larmor.frequency}
\Omega_z = -\frac{e}{2mc} B_z
\end{equation}
given by the Larmor frequency.

This means that there will be a circulating Larmor current
\begin{equation}
\label{eq.larmor.frequency.current}
I_L=Ne\frac{\Omega_z}{2\pi}= -\frac{B_z}{4\pi} N \frac{e^2}{mc}
\end{equation}
where $Ne$ is the total amount of charge on the particles ($N$
is not necessarily the number of particles). If we insert
$B_z=\Phi/( \pi R^2)$ we recover essentially equation
(\ref{eq.ring.current.due.A.elect}). This is why we called
$I_{\Phi}$ the Larmor current. Note that we derived
(\ref{eq.ring.current.due.A.elect}) under the assumption of
arbitrary charge to mass ratios $e_i/m_i$ but no interparticle
interaction. Here we need identical charge to mass ratios $e/m$
and an axially symmetric external field but can have arbitrary
interactions between the particles. The general results
(\ref{eq.ring.current.due.A.elect})   and
(\ref{eq.ring.current.due.A.SI}) for semiclassical electrons (or
electron pairs or groups) in cold metal rings thus seem fairly
reliable.

It is noteworthy that the result of equation
(\ref{eq.ring.current.due.A}) is not necessarily due to any
magnetic field affecting the particles. The flux $\Phi$ could
very well go through a smaller surface completely inside the ring
material. This means that the current in
(\ref{eq.ring.current.due.A}) is a classical Aharonov-Bohm effect
\cite{SCaharonov}. That is, an effect due to the vector
potential at zero magnetic field. By contrast the
Larmor result (\ref{eq.larmor.frequency.current}) is derived
assuming that the magnetic field penetrates the ring.

\subsection{Quantizing the electron on the circle and flux
periodicity}
The above results are purely classical. When we quantize them we
will find that physical properties must be periodic in (half?)
the flux quantum, as will now be shown show. Our previous
classical results for currents must be thought of as averages
over these quantum periods (beats). Flux quantization was
originally suggested by London \cite{BKlondon}, for a thorough
discussion see Thouless \cite{BKthouless}.

The classical Hamiltonian of an electron moving freely on
a circle of radius
$R$ threaded by a flux $\Phi$ is, according to equation
(\ref{eq.ring.hamiltonian.ext.A}),
\begin{equation}
\label{eq.Ham.free.elec.circ}
H=\frac{1}{2mR^2}\left(J +\frac{|e|}{2\pi c}\Phi \right)^2.
\end{equation}
We quantize this by letting $J \rightarrow \hat{J}=-{\rm
i}\hbar \partial /\partial \varphi$ and thus get the
Schr\"odinger equation
\begin{equation}
\label{eq.schrod.eq.free.elec.circ}
\frac{\hbar^2}{2mR^2}\left(-{\rm
i}\frac{\partial}{\partial\varphi} +n_{\phi} \right)^2
\psi(\varphi)=E\psi(\varphi),
\end{equation}
where we have used equation (\ref{eq.flux.unit}). Putting
\begin{equation}
\label{eq.gauge.transf}
\psi(\varphi)=\exp(-{\rm i}n_{\phi}\varphi)\,\psi'(\varphi)
\end{equation}
we get
\begin{equation}
\label{eq.schrod.eq.free.elec.circ.gt}
-\frac{\hbar^2}{2mR^2}\frac{\partial^2
}{\partial\varphi^2}\psi' =E \psi'
\end{equation}
for the gauge transformed wave function. It is now frequently
argued \cite{SCbyers,SCbloch} that the wave function must be single
valued and that therefore
\begin{equation}
\label{eq.single.value.cond}
\psi(\varphi+2\pi)=\psi(\varphi).
\end{equation}
Via (\ref{eq.gauge.transf}) this leads to the
physical condition
\begin{equation}
\label{eq.single.value.cond.gt}
\psi'(\varphi+2\pi)=\exp({\rm i}n_{\phi}2\pi) \psi'(\varphi)
\end{equation}
on the solutions of (\ref{eq.schrod.eq.free.elec.circ.gt}),
where the flux has been transformed away. This boundary
condition is unchanged if $n_{\phi}$ changes by unity.
This implies that physical quantities must be periodic in the
flux with period $\Phi_0$.

The above argument is not necessarily reliable, however. The
correct wave function for an electron is a spinor (in the
non-relativistic case a two component spinor). A spinor is
well known to change sign when rotated by $2\pi$. The
question is then: will the spinor rotate as the electron
travels round the circle? A free electron is known to have
conserved helicity, the projection of the spin on the
momentum. As the ring radius is large compared to atomic
dimensions the electron momentum turns slowly and it seems
reasonable that the helicity will remain conserved (as an
adiabatic invariant). This, of course, means that the spinor
must rotate with the momentum. The conclusion of all this is
that the correct condition on the spinor wave function, for a
single electron, should be
\begin{equation}
\label{eq.double.value.cond}
\psi(\varphi+4\pi)=\psi(\varphi),
\end{equation}
and thus that
\begin{equation}
\label{eq.double.value.cond.gt}
\psi'(\varphi+4\pi)=\exp({\rm i}n_{\phi}4\pi)\psi'(\varphi).
\end{equation}
This condition is unchanged whenever $n_{\phi}$ changes by
one half. I.e.\ physical quantities must be periodic in the
flux with period $\Phi_0/2$. Note that the same result is
obtained if $|e|$ in equation (\ref{eq.Ham.free.elec.circ}) is
changed to $2|e|$. The $n_{\phi}$ in
(\ref{eq.schrod.eq.free.elec.circ}) changes to $2 n_{\phi}$ and
equation (\ref{eq.single.value.cond.gt}) becomes identical to
(\ref{eq.double.value.cond.gt}).

In conclusion the observation of the $\Phi_0/2$ periodicity does
not necessarily imply electron pairs. It might be due to single
electrons going round the ring with conserved helicity. Both the
$\Phi_0$ and the $\Phi_0/2$ periodicities have been experimentally
observed \cite{SCdeaver,SCdoll,SCgough,SCwebb,SClevy,SCchandrasekhar}.

\section{Rotating superconductors and the number of
superconducting electrons}
\label{sec.london.moment}
 There is another surprising result concerning circulating
electrons that is easily explained by Larmor's theorem
(\ref{eq.larmor.frequency}). London  \cite{BKlondon} showed (see
also \cite{SCbrady,SCcabrera,SCliu}), using his phenomenological theory
of superconductivity, that a superconducting sphere that rotates
with angular velocity
$\vecOmega$ will have an induced magnetic field (Gaussian units)
\begin{equation}
\label{eq.london.mag.field}
\vecB = \frac{2 mc}{|e|} \vecOmega
\end{equation}
in its interior. Here $m$ and $e$ are the mass and charge of
the electron. This prediction has been experimentally verified
with considerable accuracy and is equally true for high
temperature and heavy fermion superconductors
\cite{SCtate,SCsanzari}. With minor modifications it is also valid for
other axially symmetric shapes of the body, for example cylinders
or rings.

\subsection{Understanding the London moment}
The London field, or `moment', (\ref{eq.london.mag.field}) can
be thought of as follows. Assume that the superconducting body
can be viewed as a system of interacting particles with the
electronic charge to mass ratio confined by an axially symmetric
external potential. When the body rotates we can transform the
equations of motion to a co-rotating system, in which it is at
rest, but in this system the particles will be affected by a
Coriolis force $-m2\vecOmega\times \vecv$. Larmor's theorem
teaches us that such  a Coriolis force is equivalent to an
external magnetic field. Magnetic fields are, however, not
allowed inside superconductors according to the Meissner effect.
To get rid of the Coriolis forces the rotation induces surface
supercurrents that produce a suitable compensating magnetic field
$\vecB$. The Lorentz force of this field is $-(|e|/c)\vecv\times
\vecB$. Provided the relation between $\vecB$ and $\vecOmega$ is
given by (\ref{eq.london.mag.field}) the two forces cancel. The
equations of motion in the rotating system are then the same, in
the interior, as if the system did not rotate. The disturbance
from the rotation on the dynamics is minimized.

The above explanation may sound compelling, but the most direct
way of understanding formula (\ref{eq.london.mag.field}) is,
in fact, much simpler. The superconducting electrons, which are
always found just inside the surface \cite{BKlondon}, are not
dragged by the positive ion lattice so when it starts to rotate
the superconducting electrons ignore this and remain in whatever
motion they prefer. This, however, means that there will be an
uncompensated motion of positive charge density on the surface
of the body. This surface charge density, $\sigma$, will, of
course, be the same  as the density of superconducting
electrons, but of opposite sign, and will produce the magnetic
field. Using this we can calculate the number, $N$, of
superconducting electrons.

\subsection{The number of superconducting electrons}
It is well known that a rotating uniform surface charge density
will produce a uniform interior magnetic field in a sphere. If
this rotating surface charge density is $\sigma$, then the total
charge $Q$ is given by
\begin{equation}
Q=N |e|= 4\pi R^2 \sigma,
\end{equation}
and the resulting magnetic field in the interior is
\begin{equation}
\label{eq.int.mag.field.surface.dens}
\vecB = \frac{2}{3}\frac{Q}{cR} \vecOmega =
\frac{8\pi}{3}\frac{\sigma R}{c} \vecOmega ,
\end{equation}
where $R$ is the radius of the sphere (relevant formulas for the
calculation can be found in Ess\'en \cite{essen89}). Putting
$Q=N|e|$ and comparing this equation with
(\ref{eq.london.mag.field}) one finds that the number $N$ must be
given by $N=3R m c^2/e^2 =3R/r_{\rm e}$. We thus find that the
relationship
\begin{equation}
\label{eq.N.re.R.sup.cond.sphere}
\frac{N r_{\rm e}}{R} =3 ,
\end{equation}
where $r_{\rm e}$ is the classical electron radius, and $N$ the
number of electrons contributing to the supercurrent,
characterizes the superconductivity on a sphere of radius $R$.

The corresponding calculation for a cylinder, long
enough for edge effects to be negligible, is elementary and gives
a similar value for $Nr_{\rm e}/\ell$, where $\ell$ is the length of the cylinder.
We will return to the crucial significance of the dimensionless
combination $N r_{\rm e}/R$ below. It is noteworthy that the
number $N$ depends only on the geometry (size) and fundamental
constants ($r_{\rm e}$). How can this be if superconductivity is
caused by some effective interaction with the lattice?

\section{Pairing and collective effects due to the Darwin-Breit
interaction}
\label{sec.darwin.breit}
We now continue the study of the semiclassical (ballistic)
electrons in the ring using the model of charged particles
constrained to move on a circle. Now we further assume that the
electrons are free particles to zeroth order and investigate how
this is affected by the first order Darwin-Breit term. The
relativistic mass-velocity correction is probably not of much
interest here.

\subsection{The Darwin-Breit term on the ring}
For the positions and velocities of equation
(\ref{eq.pos.vel.vectors.meso}) the Darwin-Breit term
(\ref{eq.V1.velocity.form}) becomes
\begin{equation}
\label{eq.darwin.breit.meso}
V_1=-\frac{e^2 }{R c^2} \sum_{i<j}^N R^2
\dot\varphi_i\dot\varphi_j
\frac{1}{4}\frac{1+3\cos(\varphi_i-\varphi_j)}{\sqrt{
2[1-\cos(\varphi_i-\varphi_j)]}} \equiv -\frac{e^2 R}{c^2}
\sum_{i<j}^N
\dot\varphi_i\dot\varphi_j V_{\varphi}(\varphi_i-\varphi_j),
\end{equation}
and the first order Lagrangian $L=T_0-V_1$, with $T_0$ given in
equation (\ref{eq.zeroth.circle.lagrangian.V0}), is
\begin{equation}
\label{eq.lagrangian.darwin.breit.meso}
L=\frac{1}{2}mR^2 \sum_{i=1}^N \dot\varphi_i^2 +
\frac{e^2 R}{c^2} \sum_{i<j}
\dot\varphi_i \dot\varphi_j V_{\varphi}(\varphi_i-\varphi_j).
\end{equation}
The nature of the
function $V_{\varphi}$ is indicated in equation
(\ref{eq.series.V.phi}) below.
If we introduce (note that the electron has charge $e=-|e|$)
\begin{equation}
\label{eq.int.vec.pot.meso}
A_i= \frac{e}{c}  \sum_{j(\neq i)}^N \dot\varphi_j
V_{\varphi}(\varphi_i-\varphi_j)
\end{equation}
we can write this
\begin{equation}
\label{eq.hamilton.darwin.breit.meso.A}
L= \sum_{i=1}^N \left( \frac{1}{2}mR^2
\dot\varphi_i^2 + \frac{e}{2c} R\dot\varphi_i A_i \right).
\end{equation}
It is easy to show that for real electrons distributed
round a (one-dimensional) ring of real atoms the Darwin-Breit
term will always be a small perturbation \cite{SCgeller}.
Individual terms in the interaction may still be large if some
pair of interparticle distances is very small. This would
correspond to pair formation and is treated in the next
subsection. In the real world of two and three dimensions the
Darwin-Breit term as a whole can become large. This means that
individually  moving particles is no longer a good first
approximation. This is shown in the following subsection.

\subsection{The one-dimensional hydrogen atom}
The Darwin-Breit term represents an interaction which is
attractive for parallel currents. For small relative velocities
of the electrons it seems possible that it could lead to bound
states (for the relative motion of the particles). Let us
investigate this. Most conduction electrons in the metal ring
will be inside the (one-dimensional) Fermi surface and they will
occur in pairs of opposite momentum with no net current. Assume
that only two electrons have unpaired momenta and move in the
same direction around the ring approximately with the Fermi
velocity. The Lagrangian of these two is then
\begin{equation}
\label{eq.Lagrangian.two.part.ring.darw}
L=  \frac{mR^2}{2}(\dot\varphi_1^2+\dot\varphi_2^2)+\frac{e^2 R}{
c^2}\dot\varphi_1 \dot\varphi_2 V_{\varphi}(\varphi_1-\varphi_2).
\end{equation}
We now make the coordinate transformation
\begin{equation}
\label{eq.two.part.ring.coord.cm.rel}
\varphi_{\mbox{\tiny C}}=\frac{1}{2}(\varphi_1 +\varphi_2),\hskip 0.5cm
\varphi=\varphi_1-\varphi_2
\end{equation}
to center of mass angle $\varphi_{\mbox{\tiny C}}$ and relative angle
$\varphi$. The inverse transformation is
\begin{equation}
\label{eq.two.part.ring.coord.cm.rel.inv}
\varphi_1=\varphi_{\mbox{\tiny C}}+\frac{1}{2}\varphi ,\hskip 0.5cm
\varphi_2=\varphi_{\mbox{\tiny C}}-\frac{1}{2}\varphi,
\end{equation}
and the Lagrangian becomes
\begin{equation}
\label{eq.Lagrangian.two.part.ring.darw.cm.rel}
L= \frac{mR^2}{2}\left(2\dot\varphi_{\mbox{\tiny C}}^2+ \frac{1}{2}
\dot\varphi^2\right) + \frac{e^2 R}{ c^2}\left(\dot\varphi_{\mbox{\tiny C}}^2
-\frac{1}{4}\dot\varphi^2\right) V_{\varphi}(\varphi).
\end{equation}
We define $J_{\mbox{\tiny C}}\equiv \partial L/\partial\dot\varphi_{\mbox{\tiny C}}$ and
$J\equiv \partial L/\partial\dot\varphi$ and get the
(exact) Hamiltonian
\begin{equation}
\label{eq.Hamiltonian.two.part.ring.darw.cm.rel}
H= J_{\mbox{\tiny C}} \dot\varphi_{\mbox{\tiny C}} +J \dot\varphi - L= \frac{1}{4}
\frac{J_{\mbox{\tiny C}}^2}{mR^2\left(1+ \frac{e^2 V_{\varphi}(\varphi)}{m c^2 R}
\right)}  +\frac{J^2}{ mR^2 \left(1 - \frac{e^2
V_{\varphi}(\varphi)}{m c^2 R}
\right)}.
\end{equation}
Clearly $\dot J_{\mbox{\tiny C}} = -\partial
H/\partial\dot\varphi_{\mbox{\tiny C}} =0$ so the center of mass (angular)
momentum $J_{\mbox{\tiny C}}$ is conserved. We put
\begin{equation}
\label{eq.def.J.F}
|J_{\mbox{\tiny C}}| \equiv 2 J_{\mbox{\tiny F}} = \mbox{const.}
\end{equation}
and expand to first order in the parameter $\frac{e^2/R}{m
c^2 }=r_{\rm e}/R$. Throwing away a constant we end up with the
following Hamiltonian for the relative motion of the electrons
\begin{equation}
\label{eq.Hamiltonian.two.part.ring.darw.rel.1}
H= \frac{J^2}{mR^2}- \frac{J_{\mbox{\tiny F}}^2 -J^2}{mR^2}
\frac{e^2}{m c^2}
\frac{V_{\varphi}(\varphi)}{R} .
\end{equation}
Consistency with our original assumptions requires that
$J^2 \ll J_{\mbox{\tiny F}}^2$ and thus we neglect the $J^2$ in
the second term. Series expansion of $V_{\varphi}$ gives
\begin{equation}
\label{eq.series.V.phi}
V_{\varphi}(\varphi)=\frac{1}{4}\frac{1+3\cos\varphi}{\sqrt{
2(1-\cos\varphi)}} = \frac{1}{|\varphi|} -
\frac{1}{3}|\varphi|+\frac{97}{5760}|\varphi|^3 +\ldots,
\end{equation}
for the angular potential energy, so  near $\varphi=0$ this is
essentially a (one-dimensional) Coulomb potential. We keep the
first term and introduce
\begin{equation}
\label{eq.def.one.dim.hydrogen}
p \equiv J/R,\;\; \mu \equiv m/2,\;\; r \equiv R\varphi ,\;\;
\ZF\equiv \frac{J_{\mbox{\tiny F}}^2/(mR^2)}{m c^2} =\frac{\EF}{m
c^2},
\end{equation}
where $\EF$ is the Fermi energy. The Hamiltonian for the relative
motion then becomes the well known Hamiltonian,
\begin{equation}
\label{eq.Hamiltonian.two.part.ring.darw.rel.2}
H= \frac{p^2}{2\mu}-  \frac{\ZF e^2}{|r|},
\end{equation}
for a (one dimensional) one electron atom with reduced mass
$\mu$ and nuclear charge $\ZF$.

The analysis above for two electrons on a circle can be done in
an almost identical way in three dimensions
\cite{essen95,essen96,essen99} and shows that the Breit
interaction can bind two electrons in their relative motion while
their center of mass moves through the metal at
the Fermi speed. The ground state energy in that case corresponds
to a temperature of $\sim 0.1$~mK. In the present one-dimensional
case all parameters are the same except the dimensionality of the
space. The one-dimensional hydrogen atom is treated in the
literature \cite{SChaines,SCrau} and the ground state energy is known
to go logarithmically to minus infinity when the dimension
approaches one. To get a finite result we must therefore take
account of the thickness, $a$, of our ring and change the potential
to
\begin{equation}
\label{eq.trunc.coulomb}
V_1(r) =-\frac{\ZF e^2}{|r|+a}.
\end{equation}
In the three dimensional case the Bohr radius of the Hamiltonian
(\ref{eq.Hamiltonian.two.part.ring.darw.rel.2}) is $a_m=
2/\ZF\approx 1.52 \cdot 10^4 \; r_s^2/a_0$, where $a_0$ is
the ordinary Bohr radius and $r_s$ the radius parameter. The
three-dimensional ground state energy is
$E_{3d}=-1/a_m^2=-\ZF^2/4$. The corresponding result for the
one-dimensional potential (\ref{eq.trunc.coulomb}) is
\cite{SChaines,SCrau}
\begin{equation}
\label{eq.trunc.coulomb.one.dim.energy}
E_{1d}=-\frac{1}{a_m^2}[2 \ln(a_m/a)]^2 .
\end{equation}
The condition for this is that $a\ll a_m$. For the gold ring
of Chandrasekhar et al. \cite{SCchandrasekhar} with $a \approx
80\,$nm  one finds that $a_m/a \approx 10^2$ using standard
values for the Fermi energy of Au. One gets similar values for
the Cu rings of Levy et al.\ \cite{SClevy}. The $1d$-condition is
thus clearly satisfied in both experiments. We thus get that the
ground state energy of the Darwin-Breit bound electron pairs
corresponds to a temperature of roughly 1 -- 2 mK. This is a bit
below the temperatures (7 mK) at which the persistent current
gold ring experiments in \cite{SCchandrasekhar} were performed,
but the order of magnitude agreement is noteworthy. In
the $10^7$ Cu-rings experiment of Levy et al.\ \cite{SClevy} the
temperature range  7 -- 400 mK was used. Physicists working
with the theory of these phenomena can certainly not ignore the
Darwin-Breit interaction and the possibility of pairing.

\subsection{When does the Darwin-Breit term become large?}
In the previous subsection we saw that the Darwin-Breit
interaction, though small, can have important qualitative effect
and lead to pairing of electrons. This effect is enhanced by
one-dimensionality because of the logarithmic divergence of
the $1/r$-interaction in one dimension. Let us now investigate
the possibility of collective effects due to this term.

We return to the Lagrangian
(\ref{eq.lagrangian.darwin.breit.meso}) and try to get the
Hamiltonian without approximation. The generalized momentum is
\begin{equation}
\label{eq.generalized.mom.ring.meso}
J_i \equiv \frac{\partial L}{\partial \dot\varphi_i} = mR^2
\dot\varphi_i +\frac{e^2R}{c^2} \sum_{j(\neq i)}^N \dot\varphi_j
V_{\varphi}(\varphi_i-\varphi_j).
\end{equation}
In order to get an exact Hamiltonian we must solve for the
$\dot\varphi_i$ in terms of the $J_i$. If we introduce the
abbreviation $V_{ij}\equiv V_{\varphi}(\varphi_i-\varphi_j)$ we
can write the $N$ equations (\ref{eq.generalized.mom.ring.meso})
\begin{equation}
\label{eq.generalized.mom.ring.meso.2}
J_i = mR^2 \left( \dot\varphi_i +\frac{r_{\rm e}}{R} \sum_{j(\neq
i)}^N  V_{ij} \dot\varphi_j \right), \hskip 0.5cm i=1,\ldots,N,
\end{equation}
($r_{\rm e}$=classical electron radius). As long as the sum here
is negligible we have $J_i \approx mR^2  \dot\varphi_i$ and
easily find an approximate Hamiltonian. For few particles, small
$N$, the sum will, in practice, never exceed the small number $N
r_{\rm e}/R$ by much, since in quantum mechanics the uncertainty
principle prevents the $V_{ij}$ from becoming to large. If,
however, $N$ is very large, the sum can still be small if the
velocities $\dot\varphi_j$ have random signs.

We see that the condition for breakdown of the approximation $J_i
\approx mR^2  \dot\varphi_i$, and thus for important collective
effects of the Darwin-Breit term, is that $N r_{\rm e}/R$ no
longer is small.  A three dimensional estimate in
\cite{essen97} shows that, in fact, magnetic energy is
minimized when
\begin{equation}
\label{eq.cond.big.collective.DB}
\frac{N r_{\rm e}}{R} \sim 1
\end{equation}
where $N$ is the number of correlated velocities. If we put
$\varepsilon_e \equiv r_{\rm e}/R$ we can write equation
(\ref{eq.generalized.mom.ring.meso.2}) in the matrix form
\begin{equation}
\label{eq.generalized.mom.ring.meso.3}
\left( \begin{array}{c} J_1 \\J_2 \\ \vdots \\ J_N \end{array}
\right) =
 mR^2 \left( \begin{array}{cccc}
1 & \varepsilon_e V_{12} & \cdots & \varepsilon_e V_{1N} \\
 \varepsilon_e V_{21} & 1 & \cdots & \varepsilon_e V_{2N} \\
\vdots & \vdots &  & \vdots \\
\varepsilon_e V_{N1} & \varepsilon_e V_{N2} & \cdots & 1
\end{array} \right)
\left( \begin{array}{c} \dot\varphi_1 \\\dot\varphi_2 \\
\vdots \\ \dot\varphi_N \end{array} \right).
\end{equation}
This shows that collective Darwin-Breit behavior is due to
"off-diagonal long range order", a concept invented by C.\ N.
Yang \cite{SCyang}. Here the concept reappears in a classical
context and arises in the Legendre transformation from the
Lagrangian, with a Darwin-Breit interaction, to the Hamiltonian.

In a real one-dimensional ring of  atoms with electrons this
cannot happen, as will be shown below. The algebra, however, is,
barring notational and other irrelevant details, the same in two
and three dimensions \cite{essen96,essen97}. We have already seen,
in equation (\ref{eq.N.re.R.sup.cond.sphere}), that this
parameter, $Nr_{\rm e}/R$, can be unity in three dimensions when
the system is superconducting. Everything thus falls nicely
into place. The Darwin-Breit term can lead to pairing of
electrons at sufficiently low temperatures. Provided one has
long range correlation of velocities it can also lead to a large
collective effect, which, in fact, seems to be superconductivity.

The condition (\ref{eq.cond.big.collective.DB}) will imply
different physics for different spatial dimension $d$. The
number $N$ of ballistic, or semiclassical, or superconducting,
or velocity-momentum correlated, electrons will be limited by
the fact that there will be at most one contributed per atom,
usually much less. Assume, for definiteness, the maximum number.
For a sample of spatial dimension $d$ and side length $R$ this
gives, very roughly,
\begin{equation}
\label{eq.max.N.R}
N_{\rm max}(d) = R^d / a_0^d,
\end{equation}
where $a_0$ is the Bohr-radius. If we put this in equation
(\ref{eq.cond.big.collective.DB}) we get
$\frac{R^d r_{\rm e}}{a_0^d R} \sim 1$ which implies that
\begin{equation}
\label{eq.max.R.of.d}
R^{d-1} \sim a_0^d / r_{\rm e}.
\end{equation}
This gives the following (minimum) sizes $R$ of superconducting
structures in spatial dimension $d$
\begin{equation}
d\rightarrow 1+ \hskip 0.5cm \Rightarrow\hskip 0.5cm  R
\rightarrow \infty, \hskip 0.9cm
\end{equation}
\begin{equation}
d=2 \hskip 0.5cm  \Rightarrow\hskip 0.5cm  R \sim a_0^2 /
r_{\rm e}
\approx 19 000\, a_0 \approx 1 \,\mu{\rm m} ,
\end{equation}
\begin{equation}
d=3 \hskip 0.5cm  \Rightarrow\hskip 0.5cm  R \sim a_0
\sqrt{a_0/r_{\rm e}}
\approx 140\, a_0 \approx 10\,{\rm nm} .
\end{equation}
As stated above, we see that $d=1$ does not permit long range
correlation. We saw that this does not mean that electron
pairs do not form. It only means that no long range collective
phenomenon (phase transition?) will be possible. Two dimensions
differ from three in that structures (samples) must be at least two
orders of magnitude larger in (linear) size.

\section{Conclusions}
The experienced theoretical physicist, should, just by looking
at formula (\ref{eq.V1.velocity.form}), see that there is trouble
with the thermodynamics ahead, since the interaction is long
range ($\sim 1/r$) and there is no natural screening mechanism
similar to that which limits the range of the Coulomb
interaction. This trouble is here identified with
superconductivity.  The main new point here, compared to the
previous investigations by the author, is the discovery that the
parameter $Nr_{\rm e}/R$, of equation
(\ref{eq.cond.big.collective.DB}), which has appeared again and
again in my study of the Darwin Hamiltonian (the exact
Hamiltonian corresponding to the Lagrangian with the Darwin-Breit
term), also miraculously appears in an estimate of the number of
superconducting electrons, equation
(\ref{eq.N.re.R.sup.cond.sphere}). This gives a direct connection
to the heart of superconductivity that was missing before. The
painful but only conclusion must be that the Darwin-Breit
interaction is {\em the} interaction between electrons that
causes superconductivity.



\begin{thebibliography}{41}
\expandafter\ifx\csname natexlab\endcsname\relax\def\natexlab#1{#1}\fi
\expandafter\ifx\csname bibnamefont\endcsname\relax
  \def\bibnamefont#1{#1}\fi
\expandafter\ifx\csname bibfnamefont\endcsname\relax
  \def\bibfnamefont#1{#1}\fi
\expandafter\ifx\csname citenamefont\endcsname\relax
  \def\citenamefont#1{#1}\fi
\expandafter\ifx\csname url\endcsname\relax
  \def\url#1{\texttt{#1}}\fi
\expandafter\ifx\csname urlprefix\endcsname\relax\def\urlprefix{URL }\fi
\providecommand{\bibinfo}[2]{#2}
\providecommand{\eprint}[2][]{\url{#2}}

\bibitem[{\citenamefont{Gutzwiller}(1990)}]{BKgutzwiller}
\bibinfo{author}{\bibfnamefont{M.~C.} \bibnamefont{Gutzwiller}},
  \emph{\bibinfo{title}{Chaos in Classical and Quantum Mechanics}}
  (\bibinfo{publisher}{Springer-Verlag}, \bibinfo{address}{New York},
  \bibinfo{year}{1990}).

\bibitem[{\citenamefont{Brack and Bhaduri}(1997)}]{BKbrack}
\bibinfo{author}{\bibfnamefont{M.}~\bibnamefont{Brack}} \bibnamefont{and}
  \bibinfo{author}{\bibfnamefont{R.~K.} \bibnamefont{Bhaduri}},
  \emph{\bibinfo{title}{Semiclassical Physics}} (\bibinfo{publisher}{Addison
  Wesley}, \bibinfo{address}{Reading, Massachusetts}, \bibinfo{year}{1997}).

\bibitem[{\citenamefont{{Darwin}}(1920)}]{darwin}
\bibinfo{author}{\bibfnamefont{C.~G.} \bibnamefont{{Darwin}}},
  \bibinfo{journal}{Phil. Mag. ser. 6.} \textbf{\bibinfo{volume}{39}},
  \bibinfo{pages}{537} (\bibinfo{year}{1920}).

\bibitem[{\citenamefont{Breit}(1929)}]{breit29}
\bibinfo{author}{\bibfnamefont{G.}~\bibnamefont{Breit}},
  \bibinfo{journal}{Phys. Rev.} \textbf{\bibinfo{volume}{34}},
  \bibinfo{pages}{553} (\bibinfo{year}{1929}).

\bibitem[{\citenamefont{Breit}(1932)}]{breit32}
\bibinfo{author}{\bibfnamefont{G.}~\bibnamefont{Breit}},
  \bibinfo{journal}{Phys. Rev.} \textbf{\bibinfo{volume}{39}},
  \bibinfo{pages}{616} (\bibinfo{year}{1932}).

\bibitem[{\citenamefont{Sucher}(1998)}]{sucher}
\bibinfo{author}{\bibfnamefont{J.}~\bibnamefont{Sucher}},
  \bibinfo{journal}{Advances in Quantum Chemistry}
  \textbf{\bibinfo{volume}{30}}, \bibinfo{pages}{433} (\bibinfo{year}{1998}).

\bibitem[{\citenamefont{Strange}(1998)}]{BKstrange}
\bibinfo{author}{\bibfnamefont{P.}~\bibnamefont{Strange}},
  \emph{\bibinfo{title}{Relativistic Quantum Mechanics, with applications in
  condensed matter and atomic physics}} (\bibinfo{publisher}{Cambridge
  University Press}, \bibinfo{address}{Cambridge}, \bibinfo{year}{1998}).

\bibitem[{\citenamefont{Cook}(1988)}]{cook}
\bibinfo{author}{\bibfnamefont{A.~H.} \bibnamefont{Cook}},
  \bibinfo{journal}{Proc. R. Soc. Lond. A} \textbf{\bibinfo{volume}{415}},
  \bibinfo{pages}{35} (\bibinfo{year}{1988}).

\bibitem[{\citenamefont{Ess{\'e}n}(1995)}]{essen95}
\bibinfo{author}{\bibfnamefont{H.}~\bibnamefont{Ess{\'e}n}},
  \bibinfo{journal}{Phys. Scr.} \textbf{\bibinfo{volume}{52}},
  \bibinfo{pages}{388} (\bibinfo{year}{1995}).

\bibitem[{\citenamefont{Ess{\'e}n}(1996)}]{essen96}
\bibinfo{author}{\bibfnamefont{H.}~\bibnamefont{Ess{\'e}n}},
  \bibinfo{journal}{Phys. Rev. E} \textbf{\bibinfo{volume}{53}},
  \bibinfo{pages}{5228} (\bibinfo{year}{1996}).

\bibitem[{\citenamefont{Ess{\'e}n}(1997)}]{essen97}
\bibinfo{author}{\bibfnamefont{H.}~\bibnamefont{Ess{\'e}n}},
  \bibinfo{journal}{Phys. Rev. E} \textbf{\bibinfo{volume}{56}},
  \bibinfo{pages}{5858} (\bibinfo{year}{1997}).

\bibitem[{\citenamefont{Ess{\'e}n}(1999)}]{essen99}
\bibinfo{author}{\bibfnamefont{H.}~\bibnamefont{Ess{\'e}n}},
  \bibinfo{journal}{J. Phys. A: Math. Gen.} \textbf{\bibinfo{volume}{32}},
  \bibinfo{pages}{2297} (\bibinfo{year}{1999}).

\bibitem[{\citenamefont{Welker}(1938)}]{SCwelker}
\bibinfo{author}{\bibfnamefont{H.}~\bibnamefont{Welker}},
  \bibinfo{journal}{Phys. Z.} \textbf{\bibinfo{volume}{39}},
  \bibinfo{pages}{920} (\bibinfo{year}{1938}).

\bibitem[{\citenamefont{Mathur et~al.}(1998)\citenamefont{Mathur, Grosche,
  Julian, Walker, Freye, Hasselwimmer, and Lonzarich}}]{SCmathur}
\bibinfo{author}{\bibfnamefont{N.~D.} \bibnamefont{Mathur}},
  \bibinfo{author}{\bibfnamefont{F.~M.} \bibnamefont{Grosche}},
  \bibinfo{author}{\bibfnamefont{S.~R.} \bibnamefont{Julian}},
  \bibinfo{author}{\bibfnamefont{I.~R.} \bibnamefont{Walker}},
  \bibinfo{author}{\bibfnamefont{D.~M.} \bibnamefont{Freye}},
  \bibinfo{author}{\bibfnamefont{R.~K.~W.} \bibnamefont{Hasselwimmer}},
  \bibnamefont{and} \bibinfo{author}{\bibfnamefont{G.~G.}
  \bibnamefont{Lonzarich}}, \bibinfo{journal}{Nature}
  \textbf{\bibinfo{volume}{394}}, \bibinfo{pages}{39} (\bibinfo{year}{1998}).

\bibitem[{\citenamefont{Capelle and Gross}(1995)}]{SCcapelle}
\bibinfo{author}{\bibfnamefont{K.}~\bibnamefont{Capelle}} \bibnamefont{and}
  \bibinfo{author}{\bibfnamefont{E.~K.~U.} \bibnamefont{Gross}},
  \bibinfo{journal}{Physics Letters A} \textbf{\bibinfo{volume}{198}},
  \bibinfo{pages}{261} (\bibinfo{year}{1995}).

\bibitem[{\citenamefont{B\"uttiker et~al.}(1983)\citenamefont{B\"uttiker, Imry,
  and Landauer}}]{SCbuttiker}
\bibinfo{author}{\bibfnamefont{M.}~\bibnamefont{B\"uttiker}},
  \bibinfo{author}{\bibfnamefont{Y.}~\bibnamefont{Imry}}, \bibnamefont{and}
  \bibinfo{author}{\bibfnamefont{R.}~\bibnamefont{Landauer}},
  \bibinfo{journal}{Phys. Lett.} \textbf{\bibinfo{volume}{96A}},
  \bibinfo{pages}{365} (\bibinfo{year}{1983}).

\bibitem[{\citenamefont{Bloch}(1970)}]{SCbloch}
\bibinfo{author}{\bibfnamefont{F.}~\bibnamefont{Bloch}},
  \bibinfo{journal}{Phys. Rev. B} \textbf{\bibinfo{volume}{2}},
  \bibinfo{pages}{109} (\bibinfo{year}{1970}).

\bibitem[{\citenamefont{Webb et~al.}(1985)\citenamefont{Webb, Washburn, Umbach,
  and Laibowitz}}]{SCwebb}
\bibinfo{author}{\bibfnamefont{R.~A.} \bibnamefont{Webb}},
  \bibinfo{author}{\bibfnamefont{S.}~\bibnamefont{Washburn}},
  \bibinfo{author}{\bibfnamefont{C.~P.} \bibnamefont{Umbach}},
  \bibnamefont{and} \bibinfo{author}{\bibfnamefont{R.~B.}
  \bibnamefont{Laibowitz}}, \bibinfo{journal}{Phys. Rev. Lett.}
  \textbf{\bibinfo{volume}{54}}, \bibinfo{pages}{2696} (\bibinfo{year}{1985}).

\bibitem[{\citenamefont{Levy et~al.}(1990)\citenamefont{Levy, Dolan, Dunsmuir,
  and Bouchiat}}]{SClevy}
\bibinfo{author}{\bibfnamefont{L.~P.} \bibnamefont{Levy}},
  \bibinfo{author}{\bibfnamefont{G.}~\bibnamefont{Dolan}},
  \bibinfo{author}{\bibfnamefont{J.}~\bibnamefont{Dunsmuir}}, \bibnamefont{and}
  \bibinfo{author}{\bibfnamefont{H.}~\bibnamefont{Bouchiat}},
  \bibinfo{journal}{Phys. Rev. Lett.} \textbf{\bibinfo{volume}{64}},
  \bibinfo{pages}{2074} (\bibinfo{year}{1990}).

\bibitem[{\citenamefont{Chandrasekhar et~al.}(1991)\citenamefont{Chandrasekhar,
  Webb, Brady, Ketchen, Gallagher, and Kleinsasser}}]{SCchandrasekhar}
\bibinfo{author}{\bibfnamefont{V.}~\bibnamefont{Chandrasekhar}},
  \bibinfo{author}{\bibfnamefont{R.~A.} \bibnamefont{Webb}},
  \bibinfo{author}{\bibfnamefont{M.~J.} \bibnamefont{Brady}},
  \bibinfo{author}{\bibfnamefont{M.~B.} \bibnamefont{Ketchen}},
  \bibinfo{author}{\bibfnamefont{W.~J.} \bibnamefont{Gallagher}},
  \bibnamefont{and}
  \bibinfo{author}{\bibfnamefont{A.}~\bibnamefont{Kleinsasser}},
  \bibinfo{journal}{Phys. Rev. Lett.} \textbf{\bibinfo{volume}{67}},
  \bibinfo{pages}{3578} (\bibinfo{year}{1991}).

\bibitem[{\citenamefont{Johnson and Kirczenow}(1998)}]{SCjohnson}
\bibinfo{author}{\bibfnamefont{B.~L.} \bibnamefont{Johnson}} \bibnamefont{and}
  \bibinfo{author}{\bibfnamefont{G.}~\bibnamefont{Kirczenow}},
  \bibinfo{journal}{Can.\ J.\ Phys.} \textbf{\bibinfo{volume}{76}},
  \bibinfo{pages}{173} (\bibinfo{year}{1998}).

\bibitem[{\citenamefont{Imry}(1997)}]{BKimry}
\bibinfo{author}{\bibfnamefont{Y.}~\bibnamefont{Imry}},
  \emph{\bibinfo{title}{Introduction to mesoscopic physics}}
  (\bibinfo{publisher}{Oxford}, \bibinfo{address}{New York},
  \bibinfo{year}{1997}).

\bibitem[{\citenamefont{Imry and Peshkin}(1997)}]{SCimry}
\bibinfo{author}{\bibfnamefont{Y.}~\bibnamefont{Imry}} \bibnamefont{and}
  \bibinfo{author}{\bibfnamefont{M.}~\bibnamefont{Peshkin}}, in
  \emph{\bibinfo{booktitle}{Electron, a centenary volume}}, edited by
  \bibinfo{editor}{\bibfnamefont{M.}~\bibnamefont{Springford}}
  (\bibinfo{publisher}{Cambrige University Press},
  \bibinfo{address}{Cambridge}, \bibinfo{year}{1997}), pp.
  \bibinfo{pages}{208--236}.

\bibitem[{\citenamefont{Landau and Lifshitz}(1976)}]{BKlandau1}
\bibinfo{author}{\bibfnamefont{L.~D.} \bibnamefont{Landau}} \bibnamefont{and}
  \bibinfo{author}{\bibfnamefont{E.~M.} \bibnamefont{Lifshitz}},
  \emph{\bibinfo{title}{Mechanics}} (\bibinfo{publisher}{Pergamon},
  \bibinfo{address}{Oxford}, \bibinfo{year}{1976}), \bibinfo{edition}{3rd} ed.

\bibitem[{\citenamefont{Geller}(1996)}]{SCgeller}
\bibinfo{author}{\bibfnamefont{M.~R.} \bibnamefont{Geller}},
  \bibinfo{journal}{Phys. Rev. B} \textbf{\bibinfo{volume}{53}},
  \bibinfo{pages}{9550} (\bibinfo{year}{1996}).

\bibitem[{\citenamefont{Ess{\'e}n}(1989)}]{essen89}
\bibinfo{author}{\bibfnamefont{H.}~\bibnamefont{Ess{\'e}n}},
  \bibinfo{journal}{Phys. Scr.} \textbf{\bibinfo{volume}{40}},
  \bibinfo{pages}{761} (\bibinfo{year}{1989}).

\bibitem[{\citenamefont{Aharonov and Bohm}(1959)}]{SCaharonov}
\bibinfo{author}{\bibfnamefont{Y.}~\bibnamefont{Aharonov}} \bibnamefont{and}
  \bibinfo{author}{\bibfnamefont{D.}~\bibnamefont{Bohm}},
  \bibinfo{journal}{Phys. Rev.} \textbf{\bibinfo{volume}{115}},
  \bibinfo{pages}{485} (\bibinfo{year}{1959}).

\bibitem[{\citenamefont{London}(1961)}]{BKlondon}
\bibinfo{author}{\bibfnamefont{F.}~\bibnamefont{London}},
  \emph{\bibinfo{title}{Superfluids, Volume 1, Macroscopic Theory of
  Superconductivity}} (\bibinfo{publisher}{Dover}, \bibinfo{address}{New York},
  \bibinfo{year}{1961}), \bibinfo{edition}{2nd} ed.

\bibitem[{\citenamefont{Thouless}(1998)}]{BKthouless}
\bibinfo{author}{\bibfnamefont{D.~J.} \bibnamefont{Thouless}},
  \emph{\bibinfo{title}{Topological quantum numbers in nonrelativistic
  physics}} (\bibinfo{publisher}{World Scientific},
  \bibinfo{address}{Singapore}, \bibinfo{year}{1998}).

\bibitem[{\citenamefont{Byers and Yang}(1961)}]{SCbyers}
\bibinfo{author}{\bibfnamefont{N.}~\bibnamefont{Byers}} \bibnamefont{and}
  \bibinfo{author}{\bibfnamefont{C.~N.} \bibnamefont{Yang}},
  \bibinfo{journal}{Phys. Rev. Lett.} \textbf{\bibinfo{volume}{7}},
  \bibinfo{pages}{46} (\bibinfo{year}{1961}).

\bibitem[{\citenamefont{Deaver and Fairbank}(1961)}]{SCdeaver}
\bibinfo{author}{\bibfnamefont{B.~S.} \bibnamefont{Deaver}, \bibfnamefont{Jr.}}
  \bibnamefont{and} \bibinfo{author}{\bibfnamefont{W.~M.}
  \bibnamefont{Fairbank}}, \bibinfo{journal}{Phys. Rev. Lett.}
  \textbf{\bibinfo{volume}{7}}, \bibinfo{pages}{43} (\bibinfo{year}{1961}).

\bibitem[{\citenamefont{Doll and N\"abauer}(1961)}]{SCdoll}
\bibinfo{author}{\bibfnamefont{R.}~\bibnamefont{Doll}} \bibnamefont{and}
  \bibinfo{author}{\bibfnamefont{M.}~\bibnamefont{N\"abauer}},
  \bibinfo{journal}{Phys. Rev. Lett.} \textbf{\bibinfo{volume}{7}},
  \bibinfo{pages}{51} (\bibinfo{year}{1961}).

\bibitem[{\citenamefont{Gough et~al.}(1987)\citenamefont{Gough, Colclough,
  Forgan, Jordan, Keene, Muirhead, Rae, Thomas, Abell, and Sutton}}]{SCgough}
\bibinfo{author}{\bibfnamefont{C.~E.} \bibnamefont{Gough}},
  \bibinfo{author}{\bibfnamefont{M.~S.} \bibnamefont{Colclough}},
  \bibinfo{author}{\bibfnamefont{E.~M.} \bibnamefont{Forgan}},
  \bibinfo{author}{\bibfnamefont{R.~G.} \bibnamefont{Jordan}},
  \bibinfo{author}{\bibfnamefont{M.}~\bibnamefont{Keene}},
  \bibinfo{author}{\bibfnamefont{C.~M.} \bibnamefont{Muirhead}},
  \bibinfo{author}{\bibfnamefont{A.~I.~M.} \bibnamefont{Rae}},
  \bibinfo{author}{\bibfnamefont{N.}~\bibnamefont{Thomas}},
  \bibinfo{author}{\bibfnamefont{J.~S.} \bibnamefont{Abell}}, \bibnamefont{and}
  \bibinfo{author}{\bibfnamefont{S.}~\bibnamefont{Sutton}},
  \bibinfo{journal}{Nature} \textbf{\bibinfo{volume}{326}},
  \bibinfo{pages}{855} (\bibinfo{year}{1987}).

\bibitem[{\citenamefont{Brady}(1982)}]{SCbrady}
\bibinfo{author}{\bibfnamefont{R.~M.} \bibnamefont{Brady}},
  \bibinfo{journal}{Journal of Low Temperature Physics}
  \textbf{\bibinfo{volume}{49}}, \bibinfo{pages}{1} (\bibinfo{year}{1982}).

\bibitem[{\citenamefont{Cabrera and Peskin}(1989)}]{SCcabrera}
\bibinfo{author}{\bibfnamefont{B.}~\bibnamefont{Cabrera}} \bibnamefont{and}
  \bibinfo{author}{\bibfnamefont{M.~E.} \bibnamefont{Peskin}},
  \bibinfo{journal}{Phys. Rev. B} \textbf{\bibinfo{volume}{39}},
  \bibinfo{pages}{6425} (\bibinfo{year}{1989}).

\bibitem[{\citenamefont{Liu}(1998)}]{SCliu}
\bibinfo{author}{\bibfnamefont{M.}~\bibnamefont{Liu}}, \bibinfo{journal}{Phys.
  Rev. Lett.} \textbf{\bibinfo{volume}{81}}, \bibinfo{pages}{3223}
  (\bibinfo{year}{1998}).

\bibitem[{\citenamefont{Tate et~al.}(1989)\citenamefont{Tate, Cabrera, Felch,
  and Anderson}}]{SCtate}
\bibinfo{author}{\bibfnamefont{J.}~\bibnamefont{Tate}},
  \bibinfo{author}{\bibfnamefont{B.}~\bibnamefont{Cabrera}},
  \bibinfo{author}{\bibfnamefont{S.}~\bibnamefont{Felch}}, \bibnamefont{and}
  \bibinfo{author}{\bibfnamefont{J.~T.} \bibnamefont{Anderson}},
  \bibinfo{journal}{Phys. Rev. Lett.} \textbf{\bibinfo{volume}{62}},
  \bibinfo{pages}{845} (\bibinfo{year}{1989}).

\bibitem[{\citenamefont{Sanzari et~al.}(1996)\citenamefont{Sanzari, Cui, and
  Karwacki}}]{SCsanzari}
\bibinfo{author}{\bibfnamefont{M.~A.} \bibnamefont{Sanzari}},
  \bibinfo{author}{\bibfnamefont{H.~L.} \bibnamefont{Cui}}, \bibnamefont{and}
  \bibinfo{author}{\bibfnamefont{F.}~\bibnamefont{Karwacki}},
  \bibinfo{journal}{Appl. Phys. Lett.} \textbf{\bibinfo{volume}{68}},
  \bibinfo{pages}{3802} (\bibinfo{year}{1996}).

\bibitem[{\citenamefont{Haines and Roberts}(1969)}]{SChaines}
\bibinfo{author}{\bibfnamefont{L.~K.} \bibnamefont{Haines}} \bibnamefont{and}
  \bibinfo{author}{\bibfnamefont{D.~H.} \bibnamefont{Roberts}},
  \bibinfo{journal}{Am. J. Phys.} \textbf{\bibinfo{volume}{37}},
  \bibinfo{pages}{1145} (\bibinfo{year}{1969}).

\bibitem[{\citenamefont{Rau}(1985)}]{SCrau}
\bibinfo{author}{\bibfnamefont{A.~R.~P.} \bibnamefont{Rau}},
  \bibinfo{journal}{Am. J. Phys.} \textbf{\bibinfo{volume}{53}},
  \bibinfo{pages}{1183} (\bibinfo{year}{1985}).

\bibitem[{\citenamefont{Yang}(1962)}]{SCyang}
\bibinfo{author}{\bibfnamefont{C.~N.} \bibnamefont{Yang}},
  \bibinfo{journal}{Rev. Mod. Phys.} \textbf{\bibinfo{volume}{34}},
  \bibinfo{pages}{694} (\bibinfo{year}{1962}).

\end{thebibliography}

\end{document}